\documentclass[a4paper]{amsart}

\usepackage[utf8]{inputenc}
\usepackage{mathbbol,mathrsfs,amsfonts,amssymb,amsthm,stmaryrd}
\usepackage{amsmath}
\usepackage[all]{xypic}
\usepackage{mathpazo}
\usepackage[mathcal]{euscript}
\usepackage{graphicx, floatflt}
\usepackage{multicol}
\usepackage{sidecap}
\usepackage{nicefrac}
\usepackage{booktabs}
\usepackage{caption,morefloats}
\usepackage[colorlinks,hyperindex]{hyperref}
\hypersetup{linkcolor=blue, citecolor=blue}

\newcommand{\m}{{\,\textrm{m} }}

\theoremstyle{theorem}

\newtheorem{thm}{Theorem}[section]

\newtheorem{prop}[thm]{Proposition}
\newtheorem{lem}[thm]{Lemma}

\newenvironment{pf}{\begin{proof}[\textrm{{\bf Proof}}]} {\end{proof}}

\begin{document}

\title[The Crush-Down Equation]{The Crush-Down Equation\\
for Non-Constant 
Velocity Profiles}

\author[Ansgar Schneider]{Ansgar Schneider}

\thanks{ansgar.schneider@uni-muenster.de}

\begin{abstract}
Bažant et al.~have proposed 
a model for a gravity-driven collapse of a tall building
that collapses after column failure in a single storey. 
Therein the collapsing building is described by three distinct
sections. The top section which consists of the part above the first 
failing storey, the middle section which is pushed from above by the top section
and consists of compacted building material, 
and the part of the building below which is still undamaged. The middle part is gaining height during the collapse, 
the lower section is loosing height. 
The resulting equation of motion is called Crush-Down Equation.

In a first approach Bažant and Verdure used a constant velocity profile for the middle 
section, namely the top section and the middle section are assumed to have the same velocity. 
In a second approach by Bažant, Le, Greening and Benson this assumption is dropped and the model is slightly modified.
However, their modifications are based on unphysical assumptions
and lead to an erroneous version of the Crush-Down Equation.

We give a detailed account  of how to implement a non-trivial velocity profile 
for the middle section and thereby derive a 
more accurate version of the Crush-Down Equation.
\end{abstract}

\maketitle


\tableofcontents
\sloppy
\newpage
\section{Introduction}
\noindent
\subsection{Background}
On the 11th of September 2001 both of the Twin Towers 
of the World Trade Center in New York City were struck by an 
aircraft. 
The North Tower collapsed approximately one and a half hour after the 
aircraft impact, the South Tower approximately one hour after the aircraft impact.
Both buildings collapsed rapidly in a top to down manner.

The National Institute of Standards and Technology released a report in 2005
whose objective was to explain how the collapse of the towers initiated \cite{NIST1}. 
However, they did not even try to explain how the collapse progressed. 

Two years later, in 2007, a  
model was proposed by Bažant and Verdure   
to describe the gravity-driven collapse of 
a tall building as a progressive floor collapse \cite{BaVe07}.
The heart of this model is the so-called Crush-Down Equation
that describes the downward movement of the crushing front 
of the collapsing building.
Another year later Bažant, Benson, Greening and Le
modified this model slightly \cite{BBGL08}.

In this paper we shall not be concerned about what can be learned 
from the model when empirical data of the actual collapsing buildings 
is taken into account. This is done elsewhere \cite{Schn17}.
This paper is solely focused on the modifications that are made 
on the Crush-Down Equation in \cite{BBGL08}.

\begin{figure}[b]
	\includegraphics[scale=0.4, angle=0]{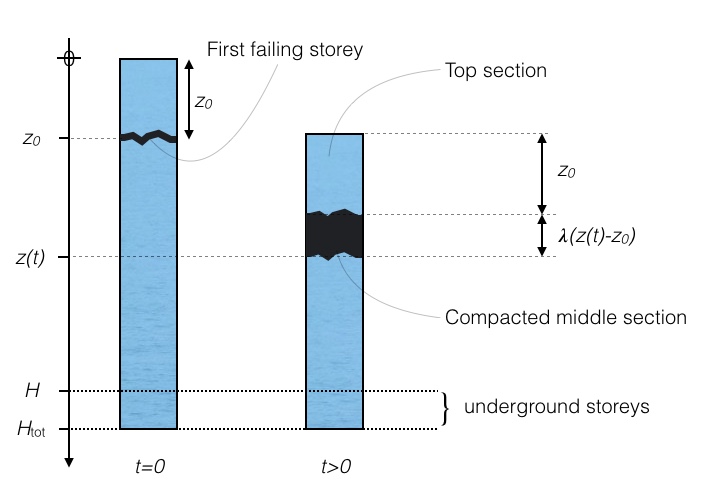}
	\caption{Schematic illustration of the gravity-driven collapse of a tall building.}
	\label{SchematicCollapse}
\end{figure}

\subsection{The Crush-Down Equation}
Let us introduce some notation and the Crush-Down Equation of \cite{BaVe07}.

Consider a tall 1-dimensional building whose roof has an elevation of $H$ over concourse level (cp.~Figure~\ref{SchematicCollapse}). 
The total height of the building including its underground storeys will be denoted by $H_{\rm tot}\ge H$.
We fix a coordinate system which is pointing 
downwards to the ground and whose origin has a fixed elevation above concourse level,
namely the elevation of the initial undestroyed tower top $H$. 

Assume one storey of the building fails due to 
some extraordinary circumstances at the position $z_0>0$ below the tower top,
i.\,e.~on the interval $[z_0,z_0+h]$ the column strength is reduced, where $h$ is the height of one storey.
The value of $z_0$ is the height of the top section of the building that will 
now decent and destroy the building below it.
The destruction of the top section itself can be neglected. An argument for this assumption 
is given in \cite[Appendix]{BBGL08}, where a  two-sided front propagation is computed.
The upward running front of destruction terminates within a fraction of a second
after having propagated a couple of centimetres only. 

Of course, we assume here that the dynamical load of the impacting top section 
is sufficiently big such that the collapse will indeed progress.
A sufficient condition for the termination of the collapse is given in \cite{BaZh02}.
The estimate therin, however, is not optimal in the sense that a weaker condition 
already suffices to arrest the collapse \cite{Schn17a}.

The crushing front will run downwards and we denote the position 
of the crushing front at time $t$ by $z(t)\ge z_0$. 
Time starts at collapse initiation, so $z(0)=z_0$.

We model the collapsing building by three distinct parts which are:
1. The initial top section of mass $m_0$ that sat above the first failing floor. 
This section keeps its height $z_0$ 
until the crushing front hits the ground. 
2. The section below the top section which is compacted from its original undamaged 
size and moving with the same velocity as the top section. The height and mass of this section 
is growing in time. 
3. The resting, still undamaged section below these two. 
The height of this section at time $t$ is $H-z(t)$. It is reducing while the collapse proceeds.

We assume that during the collapse some fraction of material is spit outwards at the crushing front.
We denote this quantity by $\kappa_{\rm out}\in [0,1]$. A value of $\kappa_{\rm out}=0$ is used 
in   \cite{BaVe07},  $\kappa_{\rm out}=0.2$ in \cite{BBGL08}
and $\kappa_{\rm out}=0.25$ in \cite{Schn17}.

Let us first assume all storeys are compacted to the same height after the crushing front has passed by.
We describe this by the so-called compaction parameter $\lambda\in (0,1)$, i.\,e.~if
$\Delta z(t):=z(t)-z_0$ is the height of the part of the building that has been crushed already, 
then $\lambda\cdot \Delta z(t)$ is the height of the compacted
section at time $t$.

The position of the roof top at time $t$ is 
$z(t) - \lambda\Delta z(t)-z_0= (1-\lambda)\Delta z(t)$, and its time derivative 
$(1-\lambda)\dot z(t)$ is the downward velocity of both the top and the middle section.
Therefore the total momentum of the falling two sections at time $t$ is
given by 
\begin{eqnarray}
\label{FirstMomentum}
p(t)= m_0\, (1-\lambda)\, \dot z(t)     +\Delta m\big(z(t)\big)\,  (1-\lambda)\, \dot z(t),
\end{eqnarray}
where
\begin{eqnarray}
\label{AggMass}
\Delta m(z):=(1-\kappa_{\rm out})\int_{z_0}^z\mu(x)\,dx
\end{eqnarray}
is the  mass of the compacted section. 
$\mu(\cdot)$ is the mass height-density   
of the undestroyed tower. We will assume that 
$\mu$ is a strictly positive, monotonously increasing, convex continuous function.
In particular, it follows that the mass function $\Delta m(\cdot)$ is also convex. 
Most of our theory can be done for general $\mu$, but 
later we shall restrict ourselves to the case of the World Trade Center.
 
Now, the equation of motion—which is called Crush-Down Equation in \cite{BaVe07}— that is
 valid until the crushing front reaches the ground is given by
\begin{eqnarray}
\label{crushdown}
\frac{d}{dt} p(t)=m_0\,g+ \Delta m\big(z(t)\big)g-F\big(z(t)\big),
\end{eqnarray}
where $F(\cdot)>0$ is the  upward resistance force
due to column buckling, and $g$ is the acceleration of gravity.
 
Other (velocity dependent) forces can be added to $F$, but we shall not be concerned about this issue here.
A discussion of the three terms that are added in \cite{BBGL08} is given in \cite[Section 1.4]{Schn17}. 
In this paper we shall discuss the question how the total momentum $p(t)$ 
changes, when two assumptions are altered. These are:
\begin{itemize}
\item[$(a)$] 
The assumption that all storeys get compacted 
to the same height when the crushing front passes by.
\item[$(b)$]
The assumption that the compacted section has a  trivial (constant)
velocity profile.
\end{itemize}
 In  \cite{BBGL08} these two points are handled in the following way:
 \begin{itemize}
\item[$(a)$]
Instead of assuming that 
every storey is compacted to the same height, it is assumed that 
every storey is compacted to the same density.
This means that instead of $\lambda\in (0,1)$ a function
$\lambda:[z_0,H_{\rm tot}]\to (0,1)$ is considered such that $\lambda$ is proportional to $\mu$.
This assumption simplifies 
the analysis significantly as we shall see, but 
it is not clear at all  that this assumption is more realistic 
than the first one. It seems reasonable to expect that during 
the collapse the lower storeys 
get compacted to a higher density than the storeys above.
A constant compaction parameter captures at least a glimpse of this aspect.

We shall consider both cases $\lambda=const$ and $\lambda\sim\mu$
in what follows.
Note that these two cases coincide 
 if $\mu(\cdot)$ is constant itself.

\item[$(b)$]
The velocity profile of the middle section is supposed to vary linearly
from the top of the middle section down to the crushing front.
However, this modification is not done accurately in 
\cite{BBGL08} for the following  reasons:
\begin{enumerate}
\item[$(i)$]
If the velocity profile is non-trivial, then conservation of 
mass implies that the density of the compacted section is also 
varying. Yet in \cite{BBGL08} it is assumed that the density is constant.
\item[$(ii)$]
The linear velocity profile of \cite{BBGL08} is assumed to vary between 
the velocity of the top section (at the top of the compacted layer) and the velocity of the 
crushing front (at the bottom of the compacted layer). This is an extremely unphysical assumption, because 
the latter velocity is bigger than 
the first one. Realistically, the velocity at the bottom of the compacted layer 
is lower than the velocity at the top.
The velocity of the crushing front should not be regarded as 
the velocity of any mass-bearing instance, but as a quantity that 
describes the change of the geometry of the crushing building.
\end{enumerate}
\end{itemize}
The formula for the total momentum that is derived from these erroneous assumptions  
is formula (2) in \cite{BBGL08}. It can be written as 
\begin{eqnarray}
\label{Boese}
p(t)= m_0\,(1-\lambda(z))\, \dot z +\Delta m(z)\,(1-\frac{1}{2}\lambda(z) )\, \dot z.
\end{eqnarray}
Note that the second velocity term $(1-\frac{1}{2}\lambda(z) )\, \dot z$ is just the mean of $\dot z$ and $(1-\lambda(z))\dot z$,
i.\,e. that's the average velocity of the linear linear velocity profile.

\begin{samepage}
\section{Crush-Down for Non-Trivial Velocity Profiles}
\label{Theorie}
\noindent
\subsection{The Set-Up}
\label{TheSetUp}
Let us denote by $y(t)$ the position of the top of the middle 
section at time $t$. If $z(t)$ is the position of the crushing front at time $t$,
 the compacted layer has an extension 
of $\Lambda(t):=z(t)-y(t)$. In particular $\Lambda(0)=0$.
We shall derive a relation between $\Lambda$ and the compaction parameter 
$\lambda:[z_0,H]\to (0,1)$. $\lambda(z(t))$ 
describes how much the storeys are compacted at the crushing front $z(t)$.
For the two cases 
$\lambda=const$ and $\lambda\sim\mu$ the
theory which we develop is manageable.
\end{samepage}
To clarify our notation we shall use the convention $\lambda(z)=\lambda_0$ if
$\lambda$ is a constant function, $\lambda_0\in\mathbb R$.
If $\lambda\sim \mu$, we will use the convention
$\lambda(z)=\lambda_0\frac{\mu(z)}{\mu_0}$, 
where  $\mu_0:=\mu(z_0)$.

By $a$ we denote the distance from $y(t)$ to some point $y(t)+a$ in the compacted section, 
$a\in [0,\Lambda(t)]$.
Let us now assume a non-trivial velocity profile $(a,t)\mapsto v(a,t)$. $v(a,t)$ is the difference of 
$\dot y(t)$ and 
the velocity of the falling part at position $y(t)+a$ at time $t$. 
So the  downward velocity of the falling section 
at  position $y(t)+a$ and time $t$ is given by $\dot y(t)+v(a,t)$. In particular
$v(0,t)=0$. If $v(a,t)=0$ for all $(a,t)$ we are in the situation as discussed before.
The assumption of a linear velocity profile as done in \cite{BBGL08}
that reaches the  velocity $\dot z(t)$ at at the crushing front 
means that the formula 
\begin{eqnarray}
\label{linearvelo}
v(a,t)= \frac{a}{\Lambda(t)}\dot\Lambda(t)
\end{eqnarray}
holds. In fact, in that case $\dot y(t)+v(\Lambda(t),t)=\dot y(t)+\dot\Lambda(t)=\dot z(t)$.
Let us now more generally assume that the velocity 
profile is of the form
\begin{eqnarray}
v(a,t)= \eta\cdot 
w\left(\frac{a}{\Lambda(t)}\right)\cdot
\dot\Lambda(t), 
\end{eqnarray}
for $\eta\in \mathbb R$ and
$w:[0,1]\to [0,\infty)$ a positive, sufficiently smooth function 
such that $w(0)=0, w(1)=1$. Realistically, $w$ is monotonously increasing
and  convex.
The prototype of such a function is $w(x)=x^\nu$ for $\nu\ge 1$.
 $\nu=1$ gives a linear velocity profile $a\mapsto v(a,t)$,
but for all $w$
we have $v(\Lambda(t),t)=\eta\dot\Lambda(t)$.

We shall not determine $w$ explicitly. However, we 
shall derive some physical restrictions on $\eta$ and $w$.
In particular, we will see that 
 $\eta\le 0$ and $w'(0)=0$.
 This excludes linear velocity profiles except the trivial one.

\subsection{Mass Conservation}
Let $m(a,t)$ be the mass of the falling section 
between the two points $y(t)$ and $y(t)+a$ at time $t$, $a\in[0,\Lambda(t)]$.
We have the following boundary condition
\begin{eqnarray}
\label{MidMass}
m(\Lambda(t),t)=\Delta m(z(t)),
\quad\text{for all } t>0.
\end{eqnarray}
Conservation of mass requires that the change of the mass $m(a,t)$ in time 
is given by the amount of mass that is moving into or out of $[y(t), y(t)+a]$ , i.\,e.
\begin{eqnarray}
\label{Massenerhaltung}
0=\partial_t m(a,t)+\partial_a m(a,t)\cdot v(a,t),
\end{eqnarray}
where $\partial_a m(a,t)=:\rho(a,t)>0$ is the height-density at the point $y(t)+a$ at time $t$.

\subsection{The Effective Compaction Parameter}
Before we state the solution of (\ref{Massenerhaltung}) let us derive the relation between $\Lambda$ and $\lambda$.
We have by mass conservation
\begin{eqnarray}
\frac{d}{dt}m(\Lambda(t),t)&=&\partial_a m(\Lambda(t),t)\dot\Lambda(t)+\partial_t m(a,t)\\\nonumber
&=&(\dot\Lambda(t)-\eta\dot\Lambda(t) )\, \rho(\Lambda(t),t). 
\end{eqnarray}
On the other hand (\ref{MidMass}) implies that
\begin{eqnarray}
\frac{d}{dt}m(\Lambda(t),t)&=&(1-\kappa_{\rm out})\, {\dot z(t)}\, \mu(z(t)).
\end{eqnarray}
In the case of the linear velocity profile of \cite{BBGL08} as stated in (\ref{linearvelo}) we have $\eta=1$.
This implies $\kappa_{\rm out}=1$ which is physically absurd. So let us now continue 
in the case  $\eta\not= 1$. 

If $\lambda(z(t))>0$ is the (not necessarily constant) compaction ratio at the crushing front $z(t)$, 
then the density $\rho$ satisfies 
$\rho(\Lambda(t),t)=(1-\kappa_{\rm out})\, \mu(z(t))/\lambda(z(t))$.
So the two above equations imply 
\begin{eqnarray}
\label{Landl}
\Lambda(t)&=&\frac{1}{1-\eta} \int_{z_0}^{z(t)} \lambda(x)\, dx.
\end{eqnarray}
Firstly, $\Lambda(t)>0$, so $\eta<1$. Secondly, if during the collapse some part of the building has been compacted, then 
it will not extend afterwards. Therefore 
$\Lambda(t)$ must be smaller than or equal to $\int_{z_0}^{z(t)} \lambda(x)\, dx$
which means the physically meaningful  range of $\eta$ is $\eta\le0$ as we have claimed.
We will restrict ourselves from now on to $\eta\le0$.

At this stage it should be noted  that the velocity at the top of 
the compacted section is
\begin{eqnarray}
\label{Effe}
\dot y& =& \dot z - \dot \Lambda
= \dot z - \frac{\lambda(z)}{1-\eta} \dot z
=\Big(1-\underbrace{\frac{\lambda(z)}{1-\eta}}\Big)\, \dot z.\\\nonumber
&&\qquad\qquad\qquad\qquad\qquad\qquad\quad\ =:\lambda^\dag(z)
\end{eqnarray}
Because of the factor 
$1/(1-\eta)$ in (\ref{Landl}) the above defined $\lambda^\dag$ 
has—during the period of the Crush-Down—the rôle of  $\lambda$ in case of the trivial velocity profile ($\eta=0$).
We call $\lambda^\dag$ the \emph{effective compaction parameter}. 

The velocity at the bottom of the crushed section 
is
\begin{eqnarray}
\dot y(t)+v(\Lambda(t),t) &=& \left(\dot z(t)-\dot \Lambda(t)\right) +\eta \dot \Lambda(t)\\\nonumber
&=&\Big(1-(1-\eta)\lambda^\dag\big(z(t)\big)\Big)\, \dot z(t).
\end{eqnarray}
As this velocity must be positive,  we find a further restriction to $\eta$.
 E.\,g.  if $\lambda^\dag(z)\ge0.18$ for all $z$,
 we have $\eta\ge 1-\nicefrac{1}{0.18}\simeq-4.6$.

\subsection{The Derived Mass Distribution}  
We will now specify $\lambda(\cdot)$ 
to be either 
$\lambda=const$ or $\lambda\sim\mu$, 
for in these cases we can easily solve
the partial differential equation (\ref{Massenerhaltung}) with the boundary condition (\ref{MidMass}).
This can be done by integrating the characteristic curves of (\ref{Massenerhaltung}). 
On the domain $\Omega:=\{ (a,t) :  a\in (0,\Lambda(t)],\, t>0\}$ 
the solution is\footnote{
Once the solution is found one can just verify it by differentiation. We leave that computation 
to the reader.}

\begin{eqnarray}
\label{MassSolution}
\\\nonumber
m(a,t)= 
\begin{cases}
\Delta m\left(z_0+\Delta z(t)\cdot 
f\left(\frac{a}{\Lambda(t)}\right)
\right), &\text{ for } \lambda=const,\\
\Delta m(z(t)) \cdot
 f\left(\frac{a}{\Lambda(t)}\right),&\text{ for }\lambda\sim \mu,
\end{cases}
\end{eqnarray}
where
\begin{eqnarray}
\label{TheFunction}
 f(x)=
\exp\left(-\int_x^1\frac{db}{b-\eta\, w(b)}\right),\quad\text{for }x\in (0,1].
\end{eqnarray}

Of course, a physically meaningful solution 
should have an extension to $a=0$ such that
$m(0,t)=0$ for all $t\ge0$.
So to proceed to the completed domain $\overline\Omega=\{ (a,t) :  a\in [0,\Lambda(t)],\, t\ge0\}$
we extend $f$ by $f(0):=0$ and $m(0,0):=0$.
We need the following  statement which we formulate for $\eta\not=0$. 
$\eta=0$ is the case of the trivial velocity profile, where $f(x)=x$.
\begin{lem}
\label{LemmaForF}
Let $\eta< 0$. 
If $w:[0,1]\to \mathbb R$ is a positive, continuously differentiable, 
monotonously increasing function, such that $w(0)=0$ and $w(1)=1$,
then
\begin{enumerate}
\item $f$ is a continuous, concave function $[0,1]\to [0,1]$, twice continuously differentiable on $(0,1]$
such that $f(x)=0, f(1)=1$.
\item $f'(1)=(1-\eta)^{-1}$.
\item $f(x)>x$ for $x\in (0,1)$
\item If $f'(0)$ exists, then $f'$ is continuous on $[0,1]$.
\item If $f'(0)$ exists, then $w'(0)=0$.
\item If $f'(0)$ exists, then $f'(0)>1$.
\item $f'(0)$ either exists or $\lim_{x\searrow0}f'(x)=+\infty.$
\item If for some $\nu>1$, $w(x)\le x^\nu$ for all $x$ in a small neighbourhood of $0$, 
then $f'(0)$ exists.
\end{enumerate}
\end{lem}
\begin{pf}
(1) The integral in (\ref{TheFunction}) goes to $+\infty$ for $x\to 0$, so 
the limit of $f(x)$ is 0, and so $f$ is continuous.   Obviously, $f(x)\in[0,1]$ and $f(1)=1$.
Now observe that $f$ satisfies the differential equation
\begin{eqnarray*}
f'(x)=\frac{f(x)}{x-\eta \,w(x)},\quad \text{for } x>0.
\end{eqnarray*}
Differentiation gives
\begin{eqnarray*}
f''(x)=\frac{f(x)\,\eta\, w'(x) }{(x-\eta\, w(x))^2}\le 0, \quad \text{for } x>0,
\end{eqnarray*}
so $f$ is concave  on $(0,1].$ As $f$ is continuous, $f$ is concave on the compact interval.
Obviously, $f''$ exists for $x>0$ and is continuous.

(2) 
$f'(1)=(1-\eta)^{-1}$ follows directly from the differential equation for $f$.

(3)
As $\eta<0$ it follows from (2) and concaveness that $f(x)>x$ for all $x\in(0,1)$.

(4)
The differential equation for $f$ can be rewritten as
\begin{eqnarray*}
f'(x)=\frac{f(x)}{x}\cdot\frac{1}{1-\eta\frac{w(x)}{x}} \quad \text{for } x>0.
\end{eqnarray*}
Therefore so if $f'(0)=\lim_{x\searrow0}\nicefrac{f(x)}{x}$ exists, then also
\begin{eqnarray*}
\lim_{x\searrow0}f'(x)=f'(0)\cdot\frac{1}{1-\eta\,w'(0)} 
\end{eqnarray*}
exists.
So $f'(0)$ and $\lim_{x\searrow0}f'(x)$ both exist. Then they must be equal, for otherwise
one could extend $f$ differentiably to $x<0$ by two different linear functions.

(5)  
If $f'(0)$ exists, then $f'$ is continuous, so 
$f'(0) = f'(0) \cdot\nicefrac{1}{(1-\eta\,w'(0))}$ which implies
$w'(0)=0$.

(6) 
$f$ is concave with $f(0)=0$ and $f(1)=1$, so $f'(0)\ge1$.
Then by (2) it follows $f'(0)>1$.

(7)
As $f$ is concave, $f'$ is monotonously decreasing,
so $f'(0)$ does not exist if and only if $\lim_{x\to 0}f'(x)=+\infty$,
because $f'$ is continuous if $f'(0)$ exists.

(8) By (7) it suffices to show that $f(x)/(x-\eta\,w(x))$
is bounded. 
To do so assume first that 
$w(x)\le x^{\nu}$ for all $x,$
let  $h(x):=1/(1-\eta\,\nicefrac{w(x)}{x})$. 
$h$ is a strictly positive, continuous 
function on the  intervall $(0,1]$ and it is bounded from above by $1.$ 
Then
\begin{eqnarray*}
\frac{f(x)}{x-\eta\,w(x)}
&=&\exp\left(-\int_x^1\frac{1}{x-\eta\,w(x)}\, dx \right)\frac{1}{x}\,h(x)\\
&\le&\exp\left(-\int_x^1\frac{1}{x-\eta\,w(x)}\, dx \right)\frac{1}{x}\cdot 1\\
&\le&\exp\left(-\int_x^1\frac{1}{x-\eta\,x^\nu}\, dx \right)\frac{1}{x}\\
&=&\left(
\frac{1-\eta}{\big(\frac{1}{x}\big)^{\nu-1}-\eta} 
\right)^{\frac{1}{\nu-1}} \frac{1}{x}\\
&=&\left(
\frac{1-\eta}{1-\eta\,x^{\nu-1}} 
\right)^{\frac{1}{\nu-1}},
\end{eqnarray*}
and this is bounded.
We leave it to the reader to modify the above estimate if $w(x)\le x^{\nu}$ 
only on a neighbourhood of 0.
\end{pf}

Realistically,  the density $\rho(a,t)=\partial_am(a,t)$ 
is a bounded positive function.
Therefore we will assume that $f'(0)$ exists which by the above Lemma excludes 
linear velocity profiles from our discussion.  
Observe that the following formulas hold
\begin{eqnarray}
\rho(a,t)=
\begin{cases}
\frac{(1-\kappa_{\rm out})\,\mu\left(z_0+\Delta z(t)\cdot 
f\left(\frac{a}{\Lambda(t)}\right)
\right)}
{\lambda_0^\dag}
\cdot f'\left(\frac{a}{\Lambda(t)}\right)
, &\text{ for } \lambda=const,\\
\frac{(1-\kappa_{\rm out})\,\mu_0}{\lambda_0^\dag}\cdot
f'\left(\frac{a}{\Lambda(t)}\right)
 &\text{ for }\lambda
\sim  \mu.
\end{cases}
\end{eqnarray}
Here we have set $\lambda_0^\dag:=\lambda^\dag(0)$.
Therefore in both cases at $a=0$:
\begin{eqnarray}
\rho(0,t)=\frac{(1-\kappa_{\rm out})\mu_0 } {\lambda_0^\dag}f'(0), 
\end{eqnarray}
which is constant in time. We shall use this property of $\rho$ to put a meaningful boundary 
condition on $\rho$ in  Appendix~\ref{TotalComp}.

For $t>0$ let us denote by $r_t$ 
ratio of the two densities $\rho(0,t)$ at the top of the compacted section 
and $\rho(\Lambda(t),t)$ at the crushing front. I.\,e. for all $t>0$:
\begin{eqnarray}
r_t= \frac{\rho(0,t)}{\rho(\Lambda(t),t)}=
\begin{cases}
\frac{\mu(z_0)}{\mu(z(t))}\frac{f'(0)}{f'(1)},\qquad\text{for }\lambda=const,\\
\frac{f'(0)}{f'(1)},\qquad\text{for }\lambda\sim\mu,
\end{cases}
\end{eqnarray}
We assume that the initial density $\mu$ 
 is continuous and increasing towards the ground.
So
 \begin{eqnarray}
f'(0)= \sigma \, f'(1),
\end{eqnarray}
where $\sigma =\inf_t  r_t\ge1$, and by the above lemma we find for $\eta<0$ that
\begin{eqnarray}
\label{EtaSigma}
\eta> 1- \sigma .
\end{eqnarray}
E.\,g.~if we assume that during 
the collapse the variation of the density 
inside the compacted layer  is below 50\,\%, i.\,e. $\sigma =1.5$,
we find that 
\begin{eqnarray}
\eta>- 0.5.
\end{eqnarray}
50\,\% should be regarded as a pretty
high value for density variations in the compacted layer. 
Note that in \cite{BBGL08} the approximation of a  constant density
in the compacted section is made. 

\subsection{Estimates}
We continue with some technical aspects 
that will be useful later. All results in this section 
are obtained by elementary techniques.

Let $f$ be given by  (\ref{TheFunction}) 
such that  $f'(0)=\sigma f'(1)$. 
The first observation is only based on concaveness.
\begin{lem}
\label{EstimateF}
We have  
\begin{eqnarray}
\frac{1}{2}\le \int_0^1f(x)\,dx\le\frac{\sqrt\sigma }{1+\sqrt\sigma }.
\end{eqnarray}
\end{lem}
\begin{pf}
If $\eta=0$, we have $f(x)=x$ and the lemma is trivially fulfilled as $\sigma =1$. 
Now, assume $\eta<0$.
As $f$ is concave the integral can 
be estimated by the unique piece-wise 
linear function $x\mapsto h_{\sigma ,\eta}(x)$ 
that has slope $f'(0)={\sigma }\cdot{(1-\eta)^{-1}}$ at $x=0$ and slope  $f'(1)=(1-\eta)^{-1}$ at $x=1$
(two linear segments). The two segments are glued together at $x=\eta(1-\sigma )^{-1}$.
We find 
\begin{eqnarray}
\int_0^1f(x)\,dx\le
\int_0^1h_{\sigma ,\eta}(x)\,dx=1-\frac{1}{2}(1-\eta)^{-1}\left(1+\eta^2\,(\sigma -1)^{-1}\right),
\end{eqnarray}
and this expression is maximal for $\eta=1-\sqrt\sigma $. Inserting this value gives the lemma.
\end{pf}
Later we shall be interested in the term $q_0(f):=2\int_0^1f(x)\, dx-1$. 
For the above discussed value of $\sigma =1.5$ we find
$q_0(f)\le 0.102$.
For $\sigma =1.3$ we have $q_0(f)\le 0.066$, and
for $\sigma =1.1$ we have $q_0(f)\le 0.024$.

To formulate the next lemma
note that if $f$ is given by (\ref{TheFunction}), then 
the inverse function $f^{-1}$ exists and is convex and monotonously increasing.
Denote by $\widetilde f:[0,1]\to \mathbb R$ the positive 
function given by
\begin{eqnarray}
\label{TheTilde}
\widetilde f(y)= -\eta\, w(f^{-1}(y)).
\end{eqnarray}
As $w$ and $f^{-1}$ are monotonously increasing, so is $\widetilde f$.
As the composition of two monotonously increasing, convex functions
is again convex, $\widetilde f$ is convex.
(As mentioned in Section~\ref{TheSetUp},  we assume $w$ to be convex.) 

Clearly, $\widetilde f(0)=\widetilde f'(0)=0$, and $\widetilde f(1)=-\eta\le \sigma -1$.

\begin{lem}
\label{KeyLemma}
We have 
$
q_0(f)=\int_0^1\widetilde f(y)\, dy.
$
\end{lem}

\begin{pf}
Recall that by (\ref{TheFunction}) $f$ 
satisfies the differential equation
\begin{eqnarray*}
f'(x)\, (x- \eta\, w(x)) =f(x). 
\end{eqnarray*}
Therefore substituting $dy=f'(x)\,dx$ gives
\begin{eqnarray*}
\int_0^1 \widetilde f(y)\,dy
&=& \int_0^1\Big( f(x)- x\,f'(x)\Big)\,dx\\\nonumber
&=& \int_0^1 f(x)\, dx - \Big[x\,f(x)\Big]_0^1+\int_0^1f(x)\,dx
\\\nonumber
&=& 2\int_0^1 f(x)\,dx -1
\end{eqnarray*}
which proves the lemma.
\end{pf}

Similar to Lemma~\ref{KeyLemma} we find
\begin{eqnarray}
\label{qIdentity}
q_1(f)=\int_0^1 y \widetilde f(y)\,dy,
\end{eqnarray}
 for $q_1(f):=\frac{1}{2} \left( 3 \int_0^1 f(x)^{2}dx-1\right)$.

The next statement is again a statement about convexity.
\begin{lem}
\label{LemmaGamma}
Define $\gamma(f)\in \mathbb R$ by $q_0(f)=\gamma\cdot q_1(f)$,
then $\gamma(f)\in[1, \frac{3}{2}]$.
\end{lem}
\begin{pf}
Let $g$ be a positive, convex, continuous function on $[0,1]$ such that
$g(0)=0, g(1)=1$. 
Then $g(y)\le y$, so $G(x):=\int_0^x g(y)\, dy\le \frac{1}{2} x^2\le \frac{1}{2}$.
$G$ is also convex and satisfies
\begin{eqnarray*}
G(x)\le G_0(x):=\begin{cases}
\frac{1}{2} x^2, &\textrm{ for } x\le x_0\\
x\cdot x_0-\frac{1}{2}x_0^2,&\textrm{ for } x> x_0,
\end{cases}
\end{eqnarray*}
where $x_0:= 1-\sqrt{1-2\,G(1)}$.
The linear part of $G_0$ is the tangent to the parabola that goes through $G(1)$ at $x=1$.

Now, $G(1)=\int_0^1g(y)\,dy$ and $G(1)-\int_0^1G(y)\,dy=\int_0^1y\, g(y)\,dy$. Therefore
\begin{eqnarray*}
\frac{\int_0^1 g(y)\,dy }{\int_0^1y\,g(y)\,dy}
&=&
\frac{G(1)}{G(1)-\int_0^1G(x)\,dx}\\
&\le& 
\frac{G(1)}{G(1)-\int_0^1G_0(x)\,dx}\\
&=&
\frac{G(1)}{\frac{2}{3}(\sqrt{1-2G(1)}-1)+G(1) (2-\frac{1}{3}\sqrt{1-2G(1)})}.\\
\end{eqnarray*}
This fraction is a monotonously increasing function of $G(1)$.
So it is maximal for $G(1)=\frac{1}{2}$ (i.\,e. $g(y)=y$),  which gives a value of $3/2$.
Then the lemma follows for $g=\widetilde f/(-\eta)$.  
\end{pf}

Let $g$ be a positive, continuous, convex function on $[0,1]$ such that $g(0)=0$.
We will only deal with $g(y)=y\,\widetilde f(y)$ and $g(y)=\widetilde f(y)$.
For $z_1> z_0$ consider the function 
\begin{eqnarray}
\label{Sharp}
g_\#:z\mapsto \theta (z-z_1) \int_{\frac{\Delta z_1}{\Delta z}}^1 g(y)\, dy,
\end{eqnarray}
where $\Delta x:= x-z_0$ and $\theta$ is the Heaviside step function which 
vanishes for negative arguments and is constant 1 for non-negative arguments.
The following result is easy: 
\begin{lem}
The function $g_\#$ is  positive, monotonously increasing, continuous and for $z\ge z_1$ it is concave. 
Moreover $g_\#'(z_1)=\frac{1}{\Delta z_1}g(1)$ (right-sided derivative),
and  for $z\to \infty$ we have  $g_\#(z)\nearrow \int_0^1g(y)\, dy$.
\end{lem}
\begin{figure}[b]
	\includegraphics[scale=0.5, angle=0]{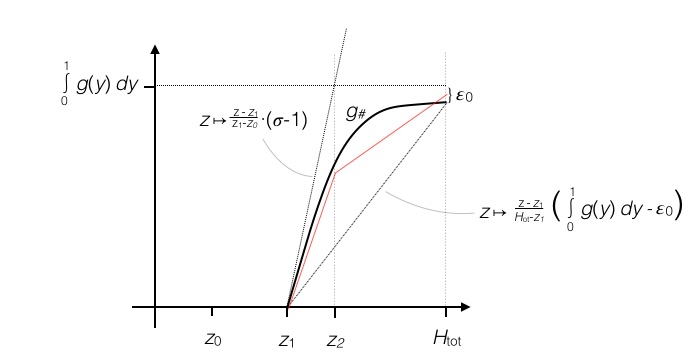}
	\caption{Approximating the concave part of the function $g_\#$.}
	\label{Approximation}
\end{figure}

We want to approximate the function $g_\#$ by a piece-wise linear function, the red function 
in Figure~\ref{Approximation}, which is the mean of the indicated upper and lower bounds. 
To make that explicit note that 
in both cases $g=\widetilde f$ and $g={\rm id}\cdot \widetilde f$
we have $g(1)=-\eta\le \sigma -1$.
Now, the total hight of the building $H_{\rm tot}$ 
should be regarded  as close to $\infty$ in the following sense:
Let $\varepsilon_0:=\int_0^\frac{\Delta z_1}{\Delta H_{\rm tot}}g(y)\, dy$.
In our main application in the next subsection we will have $z_0=46\m, z_1=110\m$ and $H_{\rm tot}=438\m$, 
so $\frac{\Delta z_1}{\Delta H_{\rm tot}}<0.164$,
and because $g$ is convex we find 
\begin{eqnarray}
\varepsilon_0<  0.164^2\cdot \int_0^1 g(y)\, dy<
0.03\cdot
\begin{cases}
q_0(f),&\text{ for }g=\widetilde f,\\
 q_1(f),&\text{ for }g={\rm id}\cdot\widetilde f.\\
\end{cases}
\end{eqnarray}
We estimate the function $g_\#$ from above by
\begin{eqnarray}
g_\#(z)&\le&g_+(z):= \theta(z-z_1) \min\left\{ \frac{z-z_1}{\Delta z_1}(\sigma-1), \int_0^1 g(y)\, dy\right\}
\end{eqnarray}
and from below by 
\begin{eqnarray}
g_\#(z)&\ge&g_-(z):= \theta(z-z_1) \frac{z-z_1}{H_{\rm tot}-z_1}\left( \int_0^1 g(y)\, dy-\varepsilon_0\right).
\end{eqnarray}
The maximal difference between $g_+$ and $g_-$ is at the gluing point $z_2$ (Figure~\ref{Approximation}).
We have $z_2=z_1+\frac{z_1-z_0}{\sigma-1}\int_0^1 g(y)\, dy$.
The  red function 
$g_{\rm red}:=\frac{1}{2}(g_++g_-)$ is the mean of $g_+$ and $g_-$.
Therefore the maximal possible error between $g_\#$ and $g_{\rm red}$ 
is smaller than
\begin{eqnarray}
\frac{1}{2}(g_+(z_2)-g_-(z_2))= \frac{1}{2}q\left (1-\frac{q-\varepsilon_0}{\sigma-1}\frac{z_1-z_0}{H_{\rm tot}-z_1}\right)
\end{eqnarray}
 at the point $z_2$. Here $q=\int_0^1 g(y)\, dy$ which is either $q=q_0(f)$ or $q=q_1(f)$.

\subsection{The Modified Crush-Down Equation}
To derive the modified Crush-Down Equation $\frac{d}{dt} p(t)= m_0g+\Delta m(z(t))g-F(z(t))$
we just need compute the aggregated momentum $p(t)$ of the top section and the middle section. We have
\begin{eqnarray}
\label{totalmomentum}
p(t)&=& m_0\, \dot y(t) +\int_0^{\Lambda(t)}\rho(a,t)\, (\dot y(t)+ v(a,t))\,da\\\nonumber 
&=&m_0\, \dot y(t) +m(\Lambda(t),t)\,\dot y(t)-   \int_0^{\Lambda(t)} \partial_t m(a,t) \,da \\\nonumber
&\stackrel{(\ref{Effe})}{=}&(m_0+\Delta m(z(t)))\, \Big(1-\lambda^\dag\big(z(t)\big)\Big)\dot z(t) -   \int_0^{\Lambda(t)} \partial_t m(a,t) \,da. 
\end{eqnarray}
So the only structurally relevant change to the momentum given by (\ref{FirstMomentum}) is the additional integral term. 
Note that $\lambda^\dag$ appears instead of $\lambda$ in  (\ref{FirstMomentum}). 

Let us discuss the additional integral term in 
two cases $\lambda=const$ and $\lambda\sim\mu$ separately. 
We start with the easier case of $\lambda\sim\mu$:
\begin{prop}
\label{PropSim}
For $\lambda\sim\mu$ we have
\begin{eqnarray}
\int_0^{\Lambda(t)} \partial_t m(a,t) \,da
&=&
q_0(f)\cdot \Delta m(z(t))\,\lambda^\dag(z(t))\, \dot z(t),
\end{eqnarray}
where $q_0(f)= 2\int_0^1f(x)\, dx-1$ as in (\ref{qIdentity}).
\end{prop}

\begin{pf}
The result follows from (\ref{MassSolution}) after substituting $db=\frac{da}{\Lambda(t)}$ and by observing that 
$$(1-\kappa_{\rm out})\,\mu(z)\,\dot z\, \Lambda=\Delta m(z)\,\lambda^\dag(z)\,\dot z$$ 
and some integration by parts.
\end{pf}
In case of the trivial velocity profile ($\eta=0$), we have $q_0(f)=0$.
For a non-trivial velocity profile we know a priori 
by Lemma~\ref{LemmaForF} that $q_0(f)\in(0,1)$.
However, by Lemma~\ref{EstimateF} 
we find a significantly lower upper bound for $q$ once we impose 
physically reasonable assumptions on the density of the middle
section. E.\,g. if we assume that the density variation in the compacted 
section is below 30\,\%, then $q_0(f)< 0.07$.
To compare this result with the modified
the Crush-Down Equation of \cite{BBGL08} 
we write the total momentum as
\begin{eqnarray}
&&p(t)=m_0\, \big(1-\lambda^\dag(z)\big)\,\dot z +\Delta m(z)\big(1-(1+q_0(f))\, \lambda^\dag(z)\big)\,\dot z.
\end{eqnarray}
This formula has exactly the same structure as (\ref{Boese}),where instead of $q_0(f)\in [0,1)$ a value of
$q=-1/2$ is used. However, $q=-1/2$ is unphysical, because it has the 
wrong sign and its absolute value is far to big.

A nice property of the formula in Proposition~\ref{PropSim} is that the time dependency (terms depending on $z$ and $\dot z$) 
and the dependency of the velocity profile (the term $q_0(f)$) split into two separate factors. 
In the case of $\lambda=const$ this is more involved, because the function $f$ appears in  (\ref{MassSolution}) in 
the  argument of the mass function $\Delta m(\cdot)$.
To state the result for the additional integral term in a manner similar to Proposition~\ref{PropSim} 
 let us introduce some short hand. For $x\in [z_0,z]$, define $\omega(x,z)\in \mathbb R$ by
\begin{eqnarray}
\Delta m'(x)=
\frac{\Delta m(z)}{\Delta z}\cdot(1+\omega(x,z)), 
\end{eqnarray}
where $\Delta z=z-z_0$ as before. 
Apparently, $\omega$ measures how much the  mass distribution $\Delta m$
fails to be linear. Note that $\omega$ does not depend on $(1-\kappa_{\rm out})$,
so $\omega$ indeed refers to the mass distribution of the undestroyed tower. 
Recall that $\mu$ is monotonously increasing, so
$x\mapsto\omega(x,z)$ is also monotonously increasing.
Note that 
by definition $\omega$ satisfies $\int_0^1\omega(z_0+\Delta z\,y,z)\, dy=0$.
Now, let
\begin{eqnarray}
Q(z,f):=\int_0^1
\omega(z_0+\Delta z\, y,z)\, \widetilde f(y)\, dy\ge0 ,
\end{eqnarray}
where $\widetilde f(y)=-\eta \,w(f^{-1}(y))$ as in (\ref{TheTilde}).
$Q(z,f)$ vanishes for a constant $\mu$.

\begin{prop} 
\label{ConsProp}
For $\lambda=const$ we have
\begin{eqnarray*}
\int_0^{\Lambda(t)} \partial_t m(a,t) \,da
&=&
\Lambda(t)\,\dot z(t)\int_0^1 \Delta m'(z_0+\Delta z(t)\, y)\, \widetilde f(y)\, dy
\\
&=&\Big(q_0(f)+Q(z(t),f)\Big) \cdot \Delta m(z(t))\,  \lambda^\dag_0\,\dot z(t).
\end{eqnarray*}
\end{prop}
\begin{pf}
The result follows from (\ref{MassSolution}) after substituting first 
$dx=\frac{da}{\Lambda(t)}$, then $dy=f'(x)\, dx$,
and then collecting the terms as in the proof of  Lemma~\ref{KeyLemma}.
\end{pf}
$Q(z(t),f)$ has a time dependency, so we need to be a little careful 
about estimating its time derivative and the resulting term that appears in the 
Crush-Down Equation.
%
To compute 
$\frac{d}{dt}p(t)$ for 
the Crush-Down Equation let us  compute the time derivative
of the additional integral term by Propostion~\ref{ConsProp}.
We have
\begin{eqnarray}
\label{ThatBloodyTimeDerivative}
&&\\\nonumber
\frac{d}{dt}\int_0^{\Lambda(t)} \partial_t m(a,t) \,da&=& 
\lambda^\dag_0\, \ddot z\,\Delta z\int_0^1 \Delta m'(z_0+\Delta z\, y)\,\widetilde f(y)\,dy\\\nonumber
&+&
\lambda^\dag_0\, \dot z^2\Big(\int_0^1\Delta m'(z_0+\Delta z\, y)\, \widetilde f(y)\,dy\\\nonumber
&&\qquad\qquad+\Delta z \int_0^1 \Delta m''(z_0+\Delta z\, y)\, y\,\widetilde f(y)\,dy\Big).
\end{eqnarray}
Let us now deal explicitly with the  mass distribution 
of the World Trade Center. In \cite{Schn17}
the following formula is used
\begin{eqnarray}
\mu(x)=\mu_0\cdot \left(1+0.43\cdot \theta(x-z_1)\frac{x-z_1}{H-z_1}\right),
\end{eqnarray}
where $\theta$ is the Heaviside step function, $z_1=110\m$, $H=417\m$ and $\mu_0=0.6\cdot10^6\,\nicefrac{\rm kg}{\!\rm m}$.
Using this explicit formula we can handle the two different 
integral terms of (\ref{ThatBloodyTimeDerivative}) using the notation of (\ref{Sharp}) :
\begin{eqnarray}
\label{Alterschwede}
&&\int_0^1 \Delta m'(z_0+\Delta z\, y)\,\widetilde f(y)\,dy\\\nonumber
&=&(1-\kappa_{\rm out})\mu_0\left( q_0(f) +0.43\frac{\Delta z}{H-z_1} ({\rm id}\cdot\widetilde f)_\#(z)
-0.43\frac{\Delta z_1}{H-z_1} \widetilde f_\#(z)\right),
\end{eqnarray}
and
\begin{eqnarray}
\label{Alteschwedin}
&&\Delta z\int_0^1 \Delta m''(z_0+\Delta z\, y)\, y\,\widetilde f(y)\,dy
=(1-\kappa_{\rm out})\mu_0\cdot0.43\frac{\Delta z}{H-z_1} ({\rm id}\cdot\widetilde f)_\#(y).
\end{eqnarray}
Note that the right-hand side of (\ref{Alteschwedin}) is the same as second summand 
in (\ref{Alterschwede}).

So far this result is exact, but as we do not have full information about the function $f$
(or the actual velocity profile of the compacted section) we need to do some approximation.  
The numerical treatment of these terms can be tackled by approximating the functions $({\rm id}\widetilde f)_\#$ and $\widetilde f_\#$ 
by $({\rm id}\widetilde f)_{\rm red}$ and $\widetilde f_{\rm red}$ as explained in the previous section.
To point out the essence of this approximation: A non-trivial velocity profile can be 
implemented in the Crush-Down Equation with three  numerical parameters 
$\sigma$, $q_0(f)$ and $q_1(f)$ (or $\gamma(f)$ from Lemma~\ref{LemmaGamma}). 
Note that in the beginning of the collapse, i.\,e.~as long as $z(t)\le z_1$ this approximation is exact 
(in this range the two cases $\lambda=const$ and $\lambda\sim\mu$ coincide).

\begin{figure}[t]
	\includegraphics[scale=0.5, angle=0]{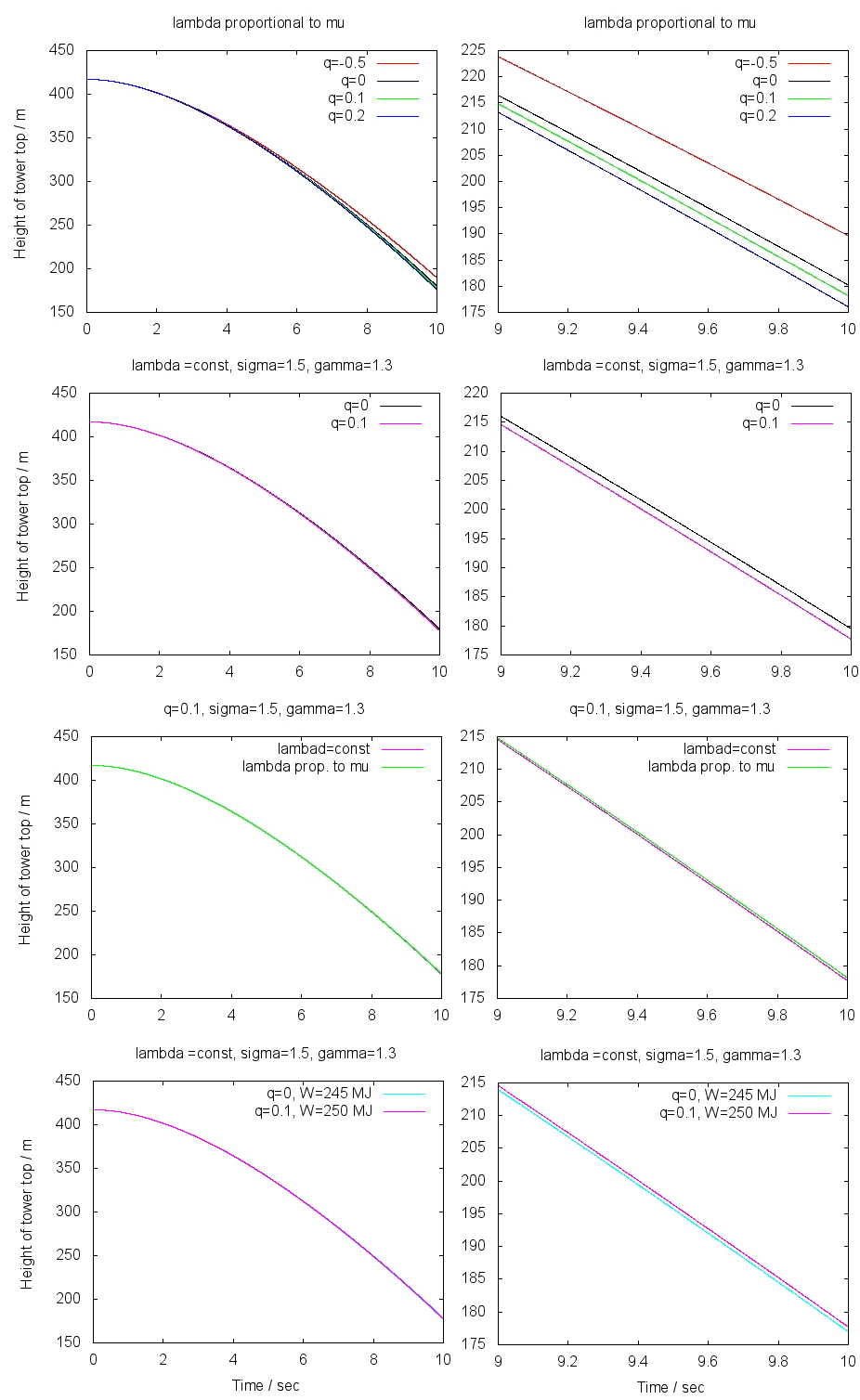}
	\caption{Comparing the solutions of the Crush-Down Equation.}
	\label{Compare}
\end{figure}

It is clear that one could give a precise estimate about error of the solution 
when $q_0(F), \sigma$ and $\gamma(f)$ vary. However, this is a tedious 
and lengthy and we will present for simplicity only a numerical treatment.
This is more illustrative, and for practical reasons this is sufficient in any case.

To obtain a numerical solution $t\mapsto z(t)$ of the Crush-Down Equation
we reformulate it as usual as a 2-dimensional differential equation 
of 1st order.
\begin{eqnarray}
\label{1sorder}
\frac{d}{dt}\left(
\begin{array}{c}
z(t)\\
v(t)
\end{array}
\right)
=
\left(
\begin{array}{c}
v(t)\\
\phi(z(t))-\psi(z(t))\,v(t)^2 
\end{array}
\right).
\end{eqnarray}
The lengthy  formulas for the 
coefficients $\phi(\cdot)$ and $\psi(\cdot)$ 
are explicitly stated in Appendix~\ref{AppNumerics}, 
where the source code for a numerical 
solution of the Crush-Down Equation is given. We use the open-source
computer algebra system Maxima (wxMaxima 16.04.0) for solving
(\ref{1sorder}) with an implementation of the Runge-Kutta algorithm.
For the computation we use the same upward resistance force $F(z)=\chi(z)F_0(z)$
as in \cite{Schn17}. $\chi:[0,H_{\rm tot}]\to [0,1]$ describes the damage of the building, 
and $F_0$ is the upward force of the undamaged columns.
Here we use a energy dissipation of 250\,MJ per storey, see 
\cite[Sec.~1.2]{Schn17} for a discussion and details. The explicit formulas are 
also stated in Appendix~\ref{AppNumerics}.
For $\kappa_{\rm out}$ we  use a value of $0.25$ as in \cite{Schn17}
and for $\lambda^\dag_0$ we use a value of $0.13$ which seems to be
reasonable in view of the considerations in Appendix~\ref{TotalComp}. 

Figure~\ref{Compare}
shows the downward movement of the roof according to different solutions of the Crush-Down Equation.
All the diagrams to the right show a zoom into the last second of the corresponding diagram to the left.
We are interested only in the first ten seconds of the solution, because this is the 
time interval for which empirical data are given in \cite{Schn17}.

In the top row the case of $\lambda\sim \mu$ for four different 
values of $q=q_0(f)$ including the unphysical value of $q=-\nicefrac{1}{2}$ is computed.
Note that in the realistic range the effect of $q$ is extremely 
small.
A value of $q=q_0(f)=0.2$ requires by Lemma~\ref{EstimateF} 
that $\sigma\ge2.25$, which means that the density of the compacted layer
varies by over 125\,\%.
Note also that the decent is faster if a non-trivial velocity profile
is assumed, whereas the erroneous assumption of \cite{BBGL08} lead to slower 
decent of the building.

The case of $\lambda=const$ is shown in the second row.
The case $q=0$ is the solution of the original unmodified Crush-Down 
equation.

The two diagrams in the third row compare  the two cases $\lambda=const$ and $\lambda\sim \mu$
explicitly. It shows the magenta solution from the second row and the green solution from the top. 
Apparently the difference is extremely small compared to any observable quantity in a realistic 
collapse scenario.

The diagrams at the bottom  compare the solution from the second row for $q=0.1$ (magenta)
with the unmodified solution $q=0$ but with a little lower energy dissipation per storey
due to column buckling (245\,MJ instead of 250\,MJ).
We find that the effect of a non-trivial velocity profile is smaller than 
a tiny variation of 2\% of the energy dissipation per storey.

\section{Discussion of Results}
\subsection{The Modified Crush-Down Equation}
In \cite{BBGL08} a modified version of the original 
Crush-Down Equation \cite{BaVe07} is used which contains some
unphysical assumptions about the compacted 
middle section of the crushing building. 

We have corrected these modification in a more realistic set-up.
The erroneous assumptions of \cite{BBGL08} lead to a wrong prediction of the movement
of the descending building. For the erroneous assumptions the predicted decent is slower
whereas under realistic assumptions the decent is slightly faster.

\subsection{Conclusion}
In \cite{Schn17} the collapse of the North Tower of the World Trade Center 
is analysed using a constant velocity profile for the compacted middle section. 
This assumption is fully justified by the presented results, because the 
uncertainty of the original parameters and the uncertainty 
of the of the measurements in \cite{Schn17} is too big to detect the effects 
of a non-trivial velocity profile.
 
 Note that the main result of \cite{Schn17} is the enormous fluctuation
 of energy dissipation during the collapse. Because the  
 building's predicted descent is faster for a non-trivial velocity profile,
 this result would be even bigger if a non-trivial velocity profile
 would be taken into account.

\begin{appendix}
\section{The Average Density of Rubble\\ and the Total Compaction Parameter}
\label{TotalComp}
\noindent
Once the crushing front reaches the ground a further compaction 
of the compacted section will take place during the subsequent Crush-Up phase.
The crushing front will start moving upwards from the ground
through the middle section. Below the upward propagating front the movement
has come to rest. Once the front reached the top of the compacted section it will start moving through the top section. 

This consideration shall  give us a boundary condition for density $\rho$ at $a=0$. 
To be consistent with our two cases $\lambda=const$ and $\lambda\sim\mu$ we  
derive two different boundary conditions.

If $\lambda\sim\mu$, then during the Crush-Down every storey is compacted to the same density 
$\rho(\Lambda(t),t)=(1-\kappa_{\rm out})\mu_0/\lambda_0$ at the crushing front.
At the top of the compacted section the density is 
$\rho(0,t)=(1-\kappa_{\rm out}){\mu_0\sigma }{\lambda_0}$ constant in time 
and we shall require that below the Crush-Up front all of the building is compacted to
$\rho(0,t)$.
This density should therefore coincide with $\mu_c$, the average density of the rubble pile. 
So our boundary condition is
\begin{eqnarray}
\mu_c\stackrel{!}{=}\rho(0,t)=(1-\kappa_{\rm out})\frac{\mu_0\, \sigma }{\lambda_0}.
\end{eqnarray}
In \cite[p.\,895]{BBGL08} a value of $\mu_c=4.1\cdot10^6\,\nicefrac{\rm kg}{\rm\!m}$ is stated 
without reference as ``\textit{typical density of rubble}" not specifying whether this means 
\emph{the} rubble of the Twin Towers or some observed rubble density of other building collapses.
Using this value we find for 
$\mu_0=0.6\cdot10^6\,\nicefrac{\rm kg}{\rm\!m}$, $\sigma =1.5,$ and $\kappa_{\rm out}=0.25$:
\begin{eqnarray}
\lambda_0 =(1-\kappa_{\rm out})\frac{\mu_0}{\mu_c}\sigma = 0.16.
\end{eqnarray}
By (\ref{EtaSigma}) this implies
\begin{eqnarray}
\lambda_0^\dag >(1-\kappa_{\rm out})\frac{\mu_0}{\mu_c}= 0.11.
\end{eqnarray}

In the case of $\lambda=const$ we assume that at the end of the Crush-Up the crushed 
building as a density distribution of\footnote{
For simplicity we ignore a discussion about what happens to the parameter 
$\kappa_{\rm out}$
when the crushing front 
is moving through the underground storeys. 
}

\begin{eqnarray}
\rho^\ddag(b){=}\frac{(1-\kappa_{\rm out})\mu\left(\frac{b}{\lambda^\ddag}\right)}{\lambda^\ddag},
\end{eqnarray}
where $b\in[0,\lambda^{\ddag}\cdot H_{\rm tot}]$ runs from the top of the collapsed building 
down to the ground. $H_{\rm tot}=H+21\m=438\m$ includes 21\m of underground storeys.
$\lambda^\ddag\in \mathbb R$ is the \emph{ total compaction parameter} defined  
by $\frac{(1-\kappa_{\rm out})\,m_{\rm tot}}{H_{\rm tot}}=\lambda^\ddag\cdot\mu_c$,
where $m_{\rm tot}=288\cdot10^6\,\rm kg$ is the total mass of the tower.\footnote{
$m_{\rm tot}= 288,000\,\rm t$ is the  value that has been 
estimated meticulously  in  \cite{Uri07}.
In \cite{BBGL08} a value of 500,000\,\rm t is stated without reference
which would give  $\lambda^\ddag=0.21$.
 }
 For $\kappa_{\rm out}=0.25$ this leads to a total compaction parameter of 
\begin{eqnarray}
\lambda^\ddag=0.11.
\end{eqnarray}
The boundary condition we impose on $\rho$ is 
\begin{eqnarray}
\rho^\ddag(z_0\lambda^\ddag)\stackrel{!}{=}\rho(0,t)=(1-\kappa_{\rm out})\frac{\mu_0\, \sigma }{\lambda_0},
\end{eqnarray}
which for the above mentioned numerical value of $\sigma=1.5$ gives 
\begin{eqnarray}
\lambda_0=\sigma \cdot\lambda^\ddag=0.18
\end{eqnarray}
and again by (\ref{EtaSigma})
\begin{eqnarray}
\lambda_0^\dag>\lambda^\ddag=0.11.
\end{eqnarray}
The inequalities become equalities for the trivial velocity profile, i.\,e. $\eta=0$.

\section{Computing   Numerical Solutions with Maxima}
\label{AppNumerics}
\noindent
The following is the Maxima source code which we have used to compute the 
solutions of Figure~\ref{Compare}. Note for the computation that
the mass density and the upward column force miss a factor $10^6$. However, 
this factor cancels out in the coefficients $\phi$ and $\psi$, so the solution is not effected 
by this simplification.

{\tiny
\texttt{
\\
/* [wxMaxima: input   start] */
\\
/* [Define the constants (lambda\_0 := \textrm{$\lambda^\dag_0$}, v\_0 = initial veolocity)] */
\\\\
g:9.8; H:417; h:3.8; z\_0:46; z\_1:110; v\_0:0;
mu\_0:0.6; lambda\_0:0.13; kappa:0.25;
\\\\
/* [The Heaviside step function] */
\\\\
theta(z):=if z<0 then 0 else 1;
\\\\
/* [The damage function and the upward resistance force] */
\\\\
chi(z):=(0.5+0.4*theta(z-z\_0-h)+0.1*theta(z-z\_0-4*h));\\
F(z):=  250/h *chi(z)*(1+theta(z-z\_1)*(6*(z-z\_1)/(H-z\_1)));
\\\\
/* [The mass density, and the  masses m\_0 := $m_0$ and Dm(z) := \textrm{$\Delta m(z)$} ] */
\\\\
mu(z):= mu\_0*(1+theta(z-z\_1)*(0.43*(z-z\_1)/(H-z\_1))) ;
\\
m\_0:mu\_0*z\_0;
\\
Dm(z):= (1-kappa)*mu\_0*(z-z\_0+ theta(z-z\_1)*0.215*(z-z\_1)\^{}2/(H-z\_1));
\\\\
/* [For $\lambda\sim\mu$ we need \%lambda(z) := $\lambda^\dag(z)$ and its derivative dlambda(z).] */\\
/* [Lambda(z) = the height of the compacted section = $\Lambda(t)$] */
\\\\
\%lambda(z):=lambda\_0*mu(z)/mu\_0;\\
dlambda(z):=lambda\_0*theta(z-z\_1)*0.43/(H-z\_1);\\
Lambda(z):=lambda\_0*Dm(z)/(mu\_0*(1-kappa));\\
\\
/* [The coefficients of the Crush-Down Equation for $\lambda\sim\mu$] */
\\\\
phi(z,q):=( (m\_0+Dm(z))*g-F(z)) / ( m\_0*(1-\%lambda(z)) + Dm(z)*(1-(1+q)*\%lambda(z)) );\\
psi(z,q):=( (1-kappa)*mu(z)*(1-(1+q)*\%lambda(z)) - (m\_0+Dm(z)*(1+q))*dlambda(z) )\\ 
\mbox{}\hspace{1cm}/ (  m\_0*(1-\%lambda(z)) + Dm(z)*(1-(1+q)*\%lambda(z))  );
\\\\
/* [Four choices of q := $q_0(f)$] */
\\\\
q\_1:-0.5;\\
q\_2:0;\\
q\_3:0.1;\\
q\_4:0.2;
\\\\
/* [Compute the solutions for $\lambda\sim\mu$] */
\\\\
time:10;\\
stepwidth:0.001;\\
solution\_1:rk([u, phi(z,q\_1)-u\^{}2*psi(z,q\_1)],[z, u],[z\_0,v\_0],[t,0,time,stepwidth])\$\\
solution\_2:rk([u, phi(z,q\_2)-u\^{}2*psi(z,q\_2)],[z, u],[z\_0,v\_0],[t,0,time,stepwidth])\$\\
solution\_3:rk([u, phi(z,q\_3)-u\^{}2*psi(z,q\_3)],[z, u],[z\_0,v\_0],[t,0,time,stepwidth])\$\\
solution\_4:rk([u, phi(z,q\_4)-u\^{}2*psi(z,q\_4)],[z, u],[z\_0,v\_0],[t,0,time,stepwidth])\$
\\\\
/* [Turn the solutions into the position of the roof (here we need the quantity Lambda(z))] */
\\\\
height\_1:makelist([solution\_1[i][1],H-(solution\_1[i][2]-Lambda(solution\_1[i][2])-z\_0)],i,1,length(solution\_1))\$\\
height\_2:makelist([solution\_2[i][1],H-(solution\_2[i][2]-Lambda(solution\_2[i][2])-z\_0)],i,1,length(solution\_2))\$\\
height\_3:makelist([solution\_3[i][1],H-(solution\_3[i][2]-Lambda(solution\_3[i][2])-z\_0)],i,1,length(solution\_3))\$\\
height\_4:makelist([solution\_4[i][1],H-(solution\_4[i][2]-Lambda(solution\_4[i][2])-z\_0)],i,1,length(solution\_4))\$
\\\\
/* [Plot the solutions for $\lambda\sim\mu$] */
\\\\
wxplot2d( 
[[discrete,height\_1],[discrete,height\_2],[discrete,height\_3],[discrete,height\_4]],\\
\mbox{}\hspace{1.5cm}[x,0,time],\\
\mbox{}\hspace{1.5cm}[style,[lines,1,red],[lines,1,black],[lines,1,green],[lines,1,blue]],\\
\mbox{}\hspace{1.5cm}[ylabel,"Height of tower top / m "], \\
\mbox{}\hspace{1.5cm}[xlabel,"Time / sec"], \\
\mbox{}\hspace{1.5cm}[title,concat("lambda proportional to mu")],\\
\mbox{}\hspace{1.5cm}[legend,concat("q=",string(q\_1)),
                    concat("q=",string(q\_2)),\\
                    \mbox{}\hspace{2.6cm}
                    concat("q=",string(q\_3)),
                    concat("q=",string(q\_4))]\\
                    )\$
\\\\
/* [For $\lambda=const$ we need the approximation of the function $g_\#$ by g\_red] */
\\\\
epsilon:0.03;\\
g\_plus(z,sigma,q):=theta(z-z\_1)*min((z-z\_1)/(z\_1-z\_0)*(sigma-1),q);\\
g\_minus(z,q):=theta(z-z\_1)*((z-z\_1)/(H+21-z\_1))*(1-epsilon)*q ;\\
g\_red(z,sigma,q):=1/2*(g\_plus(z,sigma,q)+g\_minus(z,q));\\
\\\\
/* [The coefficients of the Crush-Down Equation for $\lambda=const$] */\\
/* [gamma := $\gamma(f)$ is from Lemma~\ref{LemmaGamma}] */
\\\\
Phi(z,sigma,q,gamma):=( (m\_0+Dm(z))*g-F(z)) / ( (m\_0+ Dm(z))*(1-lambda\_0) - lambda\_0*(z-z\_0)*(1-kappa)*mu\_0 \\
\mbox{}\hspace{2cm}*(q + 0.43*(z-z\_0)/(H-z\_1)*g\_red(z,sigma,q/gamma) - 0.43*(z\_1-z\_0)/(H-z\_1)*g\_red(z,sigma,q))  );\\
Psi(z,sigma,q,gamma):=( (1-kappa)*mu(z)*(1-lambda\_0) - lambda\_0*(1-kappa)*mu\_0 \\
\mbox{}\hspace{2cm}*(q+ 2*0.43*(z-z\_0)/(H-z\_1)*g\_red(z,sigma,q/gamma) -0.43*(z\_1-z\_0)/(H-z\_1)*g\_red(z,sigma,q)) ) \\
\mbox{}\hspace{1cm}/ ( (m\_0+ Dm(z))*(1-lambda\_0) - lambda\_0*(z-z\_0)*(1-kappa)*mu\_0  \\
\mbox{}\hspace{2cm}*(q+ 0.43*(z-z\_0)/(H-z\_1)*g\_red(z,sigma,q/gamma) -0.43*(z\_1-z\_0)/(H-z\_1)*g\_red(z,sigma,q))  );
\\\\
/* [Compute the solutions for $\lambda=const$] */
\\\\
sigma:1.5;\\
gamma:1.3;\\
time:10;\\
stepwidth:0.001;\\
solution\_a:rk([u*theta(u), Phi(z,sigma,q\_2,gamma)-u\^{}2*Psi(z,sigma,q\_2,gamma)],[z, u],[z\_0,v\_0], [t,0,time,stepwidth])\$\\
solution\_b:rk([u*theta(u), Phi(z,sigma,q\_3,gamma)-u\^{}2*Psi(z,sigma,q\_3,gamma)],[z, u],[z\_0,v\_0], [t,0,time,stepwidth])\$\\
\\\\
/* [Turn the solutions into the position of the roof] */
\\\\
height\_a:makelist([solution\_a[i][1],H-(1-lambda\_0)*(solution\_a[i][2]-z\_0)],i,1,length(solution\_a))\$\\
height\_b:makelist([solution\_b[i][1],H-(1-lambda\_0)*(solution\_b[i][2]-z\_0)],i,1,length(solution\_b))\$\\
\\\\
/* [Plot the solutions for $\lambda=const$] */
\\\\
wxplot2d( [ [discrete,height\_a],[discrete,height\_b]],\\
\mbox{}\hspace{1.5cm}[x,0,time],\\
\mbox{}\hspace{1.5cm}[style,[lines,1,black],[lines,1,magenta]],\\
\mbox{}\hspace{1.5cm}[ylabel,"Height of tower top / m "], [xlabel,"Time / sec"],\\
\mbox{}\hspace{1.5cm}[title,concat("lambda =const, sigma=",string(sigma),", gamma=",string(gamma))],\\
\mbox{}\hspace{1.5cm}[legend, concat("q=",string(q\_2),""),concat("q=",string(q\_3),"")]\\
)\$
\\\\
/* [Compare the solutions for $\lambda\sim\mu$ and $\lambda=const$ ] */
\\\\
wxplot2d( [ [discrete,height\_b],[discrete,height\_3]],\\
\mbox{}\hspace{1.5cm}[x,9,time],[style,[lines,1,magenta],[lines,1,green]],\\
\mbox{}\hspace{1.5cm}[ylabel,"Height of tower top / m "], [xlabel,"Time / sec"], \\
\mbox{}\hspace{1.5cm}[title,concat("q=",string(q\_3),", sigma=",string(sigma),", gamma=",string(gamma))],\\
\mbox{}\hspace{1.5cm}[legend,concat("lambad=const"),concat("lambda prop. to mu")]\\
)\$
\\\\
/* [wxMaxima: input   end] */
}}
\end{appendix}


\begin{thebibliography}{10}

\bibitem[BaLe11]{BaLe11}
Bažant,\,Z.\,P.; Le,\,J.-L.: {\sl Why the Observed Motion History of World Trade Center Towers Is Smooth},
Journal of Engineering Mechanics, Vol. 137, No. 1, (2011),
\url{http://www.civil.northwestern.edu/people/bazant/PDFs/Papers/499.pdf}


\bibitem[BaLe16]{BaLe16}
Bažant,\,Z.\,P.; Le,\,J.-L.: {\sl Mechanics of Collapse of WTC Towers Clarified by Recent Column Buckling Tests of Korol and Sivakumaran},
Northwestern University, Report SEGIM No. 16-08c (2016),
\url{http://www.civil.northwestern.edu/people/bazant/PDFs/Papers/00-WTC-2016-buckling.pdf}



\bibitem[BaVe07]{BaVe07}
Bažant,\,Z.\,P.; Verdure, M.: {\sl  Mechanics of Progressive Collapse: 
Learning from World Trade Center and Building Demolitions},
Journal of Engineering Mechanics, Vol. 133, No. 3,  (2007),
\url{http://www.civil.northwestern.edu/people/bazant/PDFs/Papers/466.pdf}

\bibitem[BaZh02]{BaZh02}
Bažant, Z.\,P.; Zhou, Y.: {\sl Why Did the World Trade Center Collapse?—Simple Analysis},
Journal of Engineering Mechanics, Vol. 128, No. 1, (2002),
\url{http://www.civil.northwestern.edu/people/bazant/PDFs/Papers/405.pdf}




\bibitem[BLGB08]{BBGL08}
Bažant, Z.\,P; Le, J.-L.; Greening F.\,R.; Benson D.\,B.:
{\sl
What Did and Did Not Cause Collapse of World Trade Center Twin Towers in New York?}
Journal of Engineering Mechanics, Vol. 134, No. 10, (2008),,
\url{http://www.civil.northwestern.edu/people/bazant/PDFs/Papers/476.pdf}



\bibitem[NIST]{NIST1}
National Institute of Standards and Technology: {\sl NCSTAR 1: 
Federal Building and Fire Safety Investigation of the World Trade Center Disaster: 
Final Report of the National Construction Safety Team on the Collapses of the World Trade Center Tower}, (2005),
\url{https://ws680.nist.gov/publication/get_pdf.cfm?pub_id=909017}


\bibitem[Schn17a]{Schn17a}
Schneider, A.: {\sl Energy Estimates of Progressive Floor Collapses}, (2017),
preprint available at \url{http://front.math.ucdavis.edu/search?a=ansgar+schneider}

\bibitem[Schn17b]{Schn17}
Schneider, A.: {\sl The Late Jolt – Re-Examining the World Trade Center Catastrophe}, (2017),
preprint available at \url{http://front.math.ucdavis.edu/search?a=ansgar+schneider}




\bibitem[SSJ13]{JSS13}
Szulandzi\'nski,\,G.; Szamboti,\,A.; Johns,\,R.:
{\sl 
Some Misunderstandings Related to WTC Collapse Analysis},
International Journal of Protective Structures, Volume 4, Number 2,  (2013),
,
\url{http://911speakout.org/wp-content/uploads/Some-Misunderstandings-Related-to-WTC-Collapse-Analysis.pdf}


\bibitem[Uric07]{Uri07} 
Urich, G.\,H.:{\sl Analysis of the Mass and Potential Energy of World Trade Center Tower 1 },
Journal of 9/11 Studies, (2007),
\url{http://www.journalof911studies.com/letters/wtc_mass_and_energy.pdf}



\end{thebibliography}
\end{document}